\documentclass[runningheads]{llncs}
\usepackage{graphicx}
\usepackage{algorithm}
\usepackage[noend]{algpseudocode}    
\usepackage{algorithm,algpseudocode}
\usepackage{amssymb}

\usepackage{listings}
\usepackage{color}
\usepackage{times}
\usepackage{sourcecodepro}
\usepackage[T1]{fontenc}
\usepackage{listings}
\usepackage{caption}
\usepackage{courier}
\usepackage{helvet} 
\usepackage{hyperref}
\usepackage{verbatim}
\usepackage{tabularx}
\usepackage{array}

\DeclareCaptionFormat{myformat}{#3}
\lstdefinelanguage{minizinc}{
  keywords={function,if,else,endif,mod, exists},
  otherkeywords={:}
}

\usepackage{listings}
\usepackage{color}
 
\definecolor{codegreen}{rgb}{0,0.6,0}
\definecolor{codegray}{rgb}{0.5,0.5,0.5}
\definecolor{codepurple}{rgb}{0.58,0,0.82}
\definecolor{backcolour}{rgb}{0.95,0.95,0.92}
 
\lstdefinestyle{mystyle}{
    backgroundcolor=\color{backcolour},   
    commentstyle=\color{codegreen},
    keywordstyle=\color{magenta},
    numberstyle=\tiny\color{codegray},
    stringstyle=\color{codepurple},
    basicstyle=\footnotesize,
    breakatwhitespace=false,         
    breaklines=true,                 
    captionpos=b,                    
    keepspaces=true,                 
    numbers=left,                    
    numbersep=5pt,                  
    showspaces=false,                
    showstringspaces=false,
    showtabs=false,                  
    tabsize=2
}
\newcounter{lastnote}

\newcommand{\ta}{\ensuremath{\mathtt{a}}}
\newcommand{\tb}{\ensuremath{\mathtt{b}}}
\newcommand{\tc}{\ensuremath{\mathtt{c}}}

\begin{document}
\title{On Modelling 
the Avoidability of Patterns as CSP}
%
%
\author{Thorsten Ehlers\inst{1} \and
Florin Manea\inst{1} \and
Dirk Nowotka\inst{1} \and
Kamellia Reshadi\inst{1}
}
\authorrunning{T. Ehlers et al.}
%
\institute{Kiel University, Germany \\
\email{\{the,flm,dn,kre\}@informatik.uni-kiel.de}}
\maketitle              
\begin{abstract}
Solving avoidability problems in the area of string combinatorics often requires, in an initial step, the construction, via a computer program, of a very long word that does not contain any word that matches a given pattern. It is well known that this is a computationally hard task. Despite being rather straightforward that, ultimately, all such tasks can be formalized as constraints satisfaction problems, no unified approach to solving them was proposed so far, and very diverse ad-hoc methods were used. We aim to fill this gap: we show how several relevant avoidability problems can be modelled, and consequently solved, in an uniform way as constraint satisfaction problems, using the framework of MiniZinc. The main advantage of this approach is that one is now required only to formulate the avoidability problem in the MiniZinc language, and then the actual search for a solution does not have to be implemented ad-hoc, being instead carried out by a standard CSP-solver.

\end{abstract}
\section{Introduction}
The avoidability of patterns is a very well studied problem in string combinatorics. Basically, a \emph{pattern} is a string (word) that consists of \emph{terminal symbols} (e.\,g., $\ta, \tb, \tc$),  
treated as  constants, and \emph{variables} (e.\,g., $x_1, x_2, x_3$). A pattern is mapped to a word by substituting the variables by strings of terminals from an alphabet $\Sigma$.  
The avoidability problem for a pattern asks whether, for a given pattern $\alpha$ and an alphabet of terminals $\Sigma$, there exists an infinite word (also known as stream) $u$ over $\Sigma$, such that $\alpha$ does not match any finite factor of $u$. 
For more results on the avoidability of patterns, as well as applications of the results obtained in this area, see the survey \cite{currie} and the references therein, as well as the handbooks \cite{Loth97,Loth02}.

In particular, in many works on avoidability (e.g., \cite{unaryMike,unaryKre,Rao}) a substantial amount of technical lemmas were obtained with the help of computer programs. In most of the cases, these computer programs were constructing by backtracking, or simply by exhaustive exploration, very long words, over a given alphabet, which could be expressed as the morphic image of some well known infinite words, and did not contain any factor matching the given pattern. 
It is worth emphasising here that there are two main issues that make such problems hard to solve, both from the computational and the developing point of view. Firstly, checking whether a certain word matches a pattern is an NP-complete problem. So, we cannot expect that this can be done efficiently, and it usually requires a sort of exhaustive search/backtracking approach. In particular cases, this can be avoided (see, e.g., \cite{STACS2013,CiE2013,STACS2014}), but there is no general good approach to solving this problem. Then, generating the word that avoids a certain pattern is also a task which is done by exhaustive search or backtracking, which acts, in a sense, orthogonally to the task of checking whether there is a factor of the generated word matching the pattern. 

Here we propose a unifying  approach that can be used to construct long words satisfying certain avoidability properties, as described above. Exploiting the fact that, actually, we want to construct a long word satisfying a series of constraints, we will use a constraint satisfaction problem (CSP) solver to achieve this. As such, it all comes down to a rather similar solution for most avoidability problems: specify the restrictions of the problem we want to solve as constraints in the language of the solver, and then use this standardised, and usually very optimised, software to generate the long words we are looking for. 
We present here several examples, emphasising that the same ideas   can be applied for different problems, and that the resulting programs are usually much easier to read and check. Compared, e.g., to the programs we used to analyse the avoidability of patterns under permutations \cite{unaryMike,unaryKre}, this new strategy is also more efficient. Our results are described in the following.

To begin with, the CSP solver-language we use is MiniZinc. According to its authors, it is designed with the purpose of specifying constraint optimization and decision problems over integers and real numbers. The programmer specifies a model by formalizing all the constraints, without actually telling the software how to solve the problem (although the model can contain annotations which are used to guide the underlying solver). As such, the actual solution is obtained by a solver invisible to the user. MiniZinc is designed to interface easily to different backend solvers. It does this by transforming an input MiniZinc model and the input data into a FlatZinc model, which consists of variable declaration and constraint definitions as well as a definition of the objective function for optimization problems. Then a general CSP solver is used to decide whether a solution for the specified model exists, and, if yes, to find it.

We show how this approach can be used in several well-studied avoidability problems. 
Firstly, we approach the avoidability of formulas. Essentially, we are given a set of patterns and an alphabet and we want to construct a long word that does not contain any factor that matches one of the patterns in the set. We do this by specifying a MiniZinc model that defines this problem through a system of constraints, and then solving this system as a CSP. Our model allows for formulas with reversals, and it can be further constrained so that only words that are morphic image of a given standard infinite word are constructed (we used here the binary and ternary Thue-Morse words, but others can be easily used). Secondly, we show how the model can be adapted to check the avoidability of patterns in the abelian sense. Finally, we discuss the avoidability of formulas of patterns under permutations. Here the relatively simple model used in the previous cases needs to be extended with the usage of a non-trivial data file, which is, however, also automatically generated. Such data files are a standard way MiniZinc (and other modelling languages) uses to set the values of certain parameters declared in the model, based on input from the user.
We first introduce some basic combinatorics on words concepts. The description of the basics of MiniZinc can be found in the Appendix. 

\section{Preliminaries (Combinatorics on Words)}

For detailed definitions regarding combinatorics on words we refer to~\cite{Loth97},~\cite{Loth02}.

We define $\Sigma_k = \{0,\dots,k-1\}$ to be an alphabet with $k$ letters. The empty word is denoted by $\varepsilon$, $w[i]$ denotes the $i^{th}$ symbol of $w$, $|w|$ denotes the length of a word $w\in \Sigma^*_k$, and $|w|_a$ denotes the number of occurrences of the letter $a\in \Sigma_k$ in $w$. In this paper we work with \emph{terminal-free patterns}. Such a \emph{pattern} $\alpha$ is a string (word) that consists of \emph{variables} (e.\,g., $x, y, z$); we denote by $X$ the set of variables used in this paper, so $\alpha\in X^*$. A \emph{substitution} (\emph{for $\alpha$}) is a mapping $h : X \rightarrow \Sigma^+$. For every variable $x $ occurring in $\alpha$, we say that \emph{$x$ is substituted by $h(x)$}. The word  obtained by substituting every occurrence of a variable $x$ in $\alpha$ by $h(x)$ is denoted by $h(\alpha)$. 
For instance, we consider the pattern $\beta = x y y  $ and the words $u = \tb \tc \ta \tc \ta$, $v = \ta \ta \ta \ta \ta$ and have $h(\beta) = u$, for $h(x) = \tb$ and $h(y) = \tc \ta$, and $g(\beta) = v$, for $g(x) =\ta $ and $ g(y) = \ta\ta$. If there exists a substitution $h$ such that $h(\alpha)=w$, we say that $\alpha$ matches $w$. The \emph{avoidability problem} for a pattern asks whether, for a given pattern $\alpha$ and an alphabet of terminals $\Sigma$, there exists an infinite word (stream) $u$ over $\Sigma$, such that $\alpha$ does not match any finite factor of $u$. The size of the smallest alphabet (w.r.t. cardinality) over which a pattern is avoidable is the \emph{avoidability index} of that pattern.

A terminal-free pattern with functional dependencies is a term over (word) variables and function variables (where concatenation is an implicit functional constant). For example, the pattern $x \pi(y) \pi(\pi(x)) y$ has the word variables $x$ and $y$ and the function variable $\pi$. An instance of a pattern $p$ in $\Sigma_k$ is the result of substituting uniformly every variable by a word in $\Sigma_k^+$ and every function variable by a function over $\Sigma_k^*$. A pattern is avoidable in $\Sigma_k$ if there is an infinite word over $\Sigma_k$ that does not contain any instance of the pattern. Generally, we restrict the set of possible values for the function variables to morphic and antimorphic permutations of the alphabet. In \cite{unaryMike} it is shown how to find for a cube under anti-/morphic permutations $x\pi^i(x)\pi^j(x)$ all the alphabets $\Sigma_k$ over which the pattern is avoidable; as an interesting phenomenon, such a pattern is avoidable does not have an avoidability index, but rather an avoidability interval: it is unavoidable for very small and very large alphabets, and avoidable in between. A pattern with reversals is a pattern with functional dependencies, where all function variables are replaced with the mirror function, denoted here as $(\cdot)^r$.

For simplicity, from now on, terminal-free patterns are simply called patterns.

A formula $\phi$, as introduced by Cassaigne \cite{Cas}, over the set $X$ of variables is a finite set of patterns over $X$. A formula is avoidable in an alphabet $\Sigma_k$ if there exists an infinite word over $\Sigma_k$ that avoids simultaneously all the patterns in the formula. 
Cassaigne showed that every formula corresponds in a natural way to a pattern with the same avoidability index (see \cite{Cas} for details). Therefore, formulas can be seen as a natural generalization of patterns in the context of avoidability. Naturally, the notion of formulas can be also used for patterns with functional dependencies, e.g., patterns with reversals.
For example, the set $\{ xx, xyyzx^r, xyy^rx \}$ is a formula. In order to show that this formula is avoidable, an infinite word avoiding each of the patterns $xx, xyyzx^r, xyy^rx $ should be constructed. 

For a word $w$ over an alphabet $\Sigma_k$, the Parikh vector of $w$ is an array $\Psi_w$ indexed by the letters of $\Sigma_k$, such that $\Psi_w[a]=|w|_a$. Two words $u$ and $v$ are Abelian equivalent, denoted by $u \sim_{a} v$, if $v$ and $u$ have the same Parikh vector. For instance, $11122 \sim_{a} 12121, 31213 \sim_{a} 31312.$ In other words, $u\sim_a v$ means that $v$ is a jumbled version of $u$. A word $w \in \Sigma_k^{\star}$ realizes (or matches) in the Abelian sense the pattern $\alpha \in X^*$ if there are $u_1, \dots , u_{|P|} \in \Sigma_k^{+}$ such that $w = u_1 \dots u_{|\alpha|}$ and for all $i, j$ we have that $u_i\sim_{a} \alpha_j$ if and only if $\alpha[i]=\alpha[j]$. For instance, $121 121$ realizes the pattern $xx$ in the Abelian sense. 

We conclude by defining two words which are important in the study of avoidability of patterns. The infinite Thue-Morse word $t$ is defined as 
$t = \lim_{n \to \infty} \phi_t^n(0),$ 
for the morphism $\phi_t : \Sigma_2^* \to \Sigma_2^*$ where $\phi_t(0)=01$ and $\phi_t(1)=10$. It is well-known (see \cite{Loth97}) that the word $t$ avoids the patterns $xxx$ (cubes) and $xyxyx$ (overlaps).

The infinite ternary Thue word $h$ (also called sometimes the Hall word) is defined as
$h = \lim_{n \to \infty} \phi_h^n(0),$ 
for the morphism $\phi_h : \Sigma_3^* \to \Sigma_3^*$ where $\phi_h(0)=012$, $\phi_h(1)=02$ and $\phi_h(2)=1$. 
The infinite word $h$ avoids the pattern $xx$ (squares).

An archive containing the code for all the MiniZinc models we describe in the following, as well as the programs used to generate their input files and the checkers, is available at:
\href{https://media.informatik.uni-kiel.de/zs/AvoidabilityUsingMinizinc.zip}{https://media.informatik.uni-kiel.de/zs/AvoidabilityUsingMinizinc.zip}.

\section{Checking the Avoidability of Various Types of Patterns}
\subsubsection{Checking the Avoidability of Formulas with Reversal.} As announced in the introduction, we are interested in constructing words over a given alphabet $\Sigma_\ell$ that avoid a set of patterns with reversals, i.e., a formula with reversals. The input of our MiniZinc model is taken from a data file {\tt Input.dzn} in the following form. We are given the size $\ell$ of the alphabet $\Sigma_\ell$, as the parameter {\tt sigma} which is set in the data file as a positive integer. We are also given the length of the word we want to construct as the parameter {\tt wordLength}. Then we are given the number of patterns in the formula, as the parameter {\tt nrPatterns}, as well as the maximum length of a pattern {\tt maxPatternLength} and the maximum number of variables occurring in one of the patterns in the formula, {\tt maxNrVars}. Finally, we are given the patterns, in an array {\tt patterns}. A pattern with $k$ variables $x_1,\ldots,x_k$ is encoded as a word over $\{1,2,\ldots,k\}\cup \{-1,-2,\ldots,-k\}$, by replacing all the occurrences of $x_i$ with $i$ and all the occurrences of $x_i^r$ by $-i$ (for all $i\in \Sigma_k$). Connected to the formula, we are also given an array {\tt nrVarsInPattern}, which contains the number of variables occurring in each of the patterns in the formula. 

The data file is generated easily by a Java program {\tt InputGenerator.java}, which gets as input {\tt sigma}, {\tt wordLength}, and the formula written as a string of variables. Formally, the most important parameters that we send in the list of arguments of this program are  {\tt sigma, wordLength, dataFileName, pattern1, \ldots}. Alongside we need to send several other parameters {\tt t} or {\tt h}, which specifies that the generated word should be the image of the binary, respectively, ternary Thue-Morse word, {\tt morphicWordLength} which specifies the length of the prefix of {\tt t/h} that we will map to the generated word, {\tt morphicWordImagesLengths} which specifies the lengths of images of the letters  of {\tt t} or {\tt h}. These arguments will be explained in more details in Section \ref{morphic-words}. The formal of the call of this program is given in Listing 1. 
\begin{lstlisting}[caption=How to Generate Input.dzn]
java InputGenerator t/h morphicWordLength morphicWordImagesLengths sigma inputFileName pattern      
Example arguments: t 10 3 2 5 input x1x2x2x1r
\end{lstlisting}

We can now proceed and describe how the MiniZinc model was designed in order to include all the constraints fulfilled by the infinite word we want to construct. 
The general idea is the following. We want to construct a word {\tt word} that does not contain any image of the patterns in the formula. 
\begin{lstlisting}[caption=The word we construct]
array[1..wordLength] of var 1..sigma: word;
\end{lstlisting}

Therefore, we will have a set of constraints that specify that for a given pattern {\tt p} the word we construct does not contain any instance of {\tt p}. Then this set of constraints is used for all the patterns of the formula. Essentially, this is defined as in the Listing~3.
\begin{lstlisting}[caption=Dealing with formulas]
forall(p in 1..nrPatterns)( %Constraints for the pattern patterns[p]. )
\end{lstlisting}

Now, we are ready to define the constraints for the pattern {\tt patterns[p]}. We do not want any instance of this pattern to occur in the string {\tt word} we construct. So, for each position {\tt start} of the word we do not want an instance of {\tt patterns[p]} to occur starting there. To this end, we specify that for each possible assignment of the lengths of the variables occurring in {\tt patterns[p]}, the word {\tt word[start..]}  does not start with an instance of {\tt patterns[p]} under a substitution of the variables corresponding to the respective length assignment. 

More details are needed here. Firstly, we explain how we generate all the possible length assignments. It would suffice to have {\tt nrVarsInPattern[p]} stacked loops, assigning to the length of each variable values between $1$ and {\tt wordLength}. However, it is not possible to define in MiniZinc such a structure that contains a variable number of stacked loops. Therefore, a new strategy to implement this general idea is needed. For that, we will have for each assignment a {\tt label}, a variable integer. The variable {\tt label} ranges from $1$ to {\tt wordLength}$^{\mbox{\tt  nrVarsInPattern[p]}}$. Now, the length of the variable $x_i$ is encoded in {\tt label}  using the formula
\[|x_i|=\left\lceil \frac{\mbox{\tt label}}{\mbox{\tt wordLength}^{{\mbox{\tt nrVarsInPattern[p] -i}}}} \right\rceil \bmod \mbox{\tt wordLength} .\]
If this formula gives us that the length of $x_i$ is $0$, then we set the length of $x_i$ to be {\tt wordLength}. This is implemented using the function in Listing~4, where {\tt var} is the variable whose length we want to compute.

\begin{lstlisting}[caption=Computing the length of variable {\tt var} encoded by {\tt label}]
function int: length(int: var, int: wordLength, int: nrVars, int: label) =
  if (ceil(label / pow(wordLength,nrVars - var) mod wordLength == 0) then
    wordLength
  else ceil(label / pow(wordLength,nrVars - var) mod wordLength
  endif;
\end{lstlisting}

Secondly, for the lengths of the variables occurring in {\tt patterns[p]}, given by the label {\tt i}, we can compute the length of its image under a substitution consistent with those lengths. This is obtained using the MiniZinc code from Listing~5, taking into account that the patterns shorter than {\tt maxPatternLength} are padded with $0$s, up to the respective length. 

\begin{lstlisting}[caption=Length of the image of {\tt pattern[p]}]
sum(k in 1..maxPatternLength where patterns[p, k]!=0)  
		(length(abs(patterns[p, k]), wordLength, nrVarsInPattern[p], i)) - 1 
\end{lstlisting}

Once the length of the image of {\tt patterns[p]} is computed, we have two cases: this image fits in the suffix of length {\tt wordLength - start + 1} of {\tt word} or not. If not, then there is no image of {\tt patterns[p]} with the respective lengths of the variable occurring at {\tt start} in {\tt word}, so not more constraints are needed. If yes, we need to add more constraints. The idea is that looking at the string occurring at position {\tt start} in {\tt word}, whose length equals the length of the image {\tt patterns[p]}, we are able to identify its factors that correspond to the image of each occurrence of each variable. In our constraints, we ask that there are at least two such factors, that, by length reasons, should correspond to the same variable, and which are not identical. That is, we require that the respective string, occurring at position {\tt start} in {\tt word}, whose length equals the length of the image {\tt patterns[p]}, cannot be obtained by a consistent assignment of the variables of {\tt patterns[p]}, which also respects the computed lengths for the variables. This is done by the code in Listing~6. 

 \begin{lstlisting}[caption=The core constraints for {\tt pattern[p]} and label {\tt i}]
 exists(varOcc in 1..maxPatternLength where patterns[p,varOcc] != 0)(
  exists(nextOcc in (varOcc + 1)..maxPatternLength where
                 abs(patterns[p,varOcc]) == abs(patterns[p, nextOcc]))(
      let { var int: varLength = length(abs(patterns[p,varOcc]), wordLength,nrVarsInPattern[p], i),
      } in 
      exists(l in 1..varLength)(
        let { var int: occInWord = start + 
              (if varOcc > 1 then
                  (sum(k in 1..(varOcc - 1))
                  (length(abs(patterns[p, k]), wordLength, nrVarsInPattern[p], i))) 
               else 0 endif),
          var int: nextOccInWord = occInWord +
                sum(k in varOcc..(nextOcc - 1)) 
                     (length(abs(patterns[p, k]),wordLength,nrVarsInPattern[p], i)),
          var int: posFirst = occInWord + l - 1,            
          var int: posSecond =
          (if (patterns[p, varOcc] == patterns[p, nextOcc]) then (nextOccInWord+l-1)
             else (nextOccInWord + varLength - l) endif)
        } in
        word[posFirst] != word[posSecond] )))
\end{lstlisting}

Basically, in the above Listing we ask for the existence of an occurrence {\tt varOcc} of a variable $x$ in {\tt patterns[p]}, such that $x$ occurs again at least on more time on position {\tt nextOcc} of {\tt patterns[p]}. For the respective variable we denote by {\tt varLength} the length of its image in the assignment defined by the label {\tt i}. Now, we can compute the positions {\tt occInWord} and {\tt nextOccInWord}, respectively, that correspond to the positions where the images of the variable occurring on positions {\tt varOcc} and {\tt nextOcc} of {\tt patterns[p]}, respectively, occur in {\tt word}. The constraints we specify ask for the existence of a position {\tt l} such that the {\tt l$^{th}$} symbol of the first image of the variable (the one starting on {\tt occInWord}), which is found on position {\tt posFirst} in {\tt word}, is different from the {\tt l$^{th}$} symbol of the second image of the variable (the one starting on {\tt nextOccInWord}), which is found on position {\tt posSecond} in {\tt word}. A particularity of the code is that one has to take into account if the occurrence of the variable $x$ on {\tt nextOcc} is really $x$ or its mirror image $x^r$ when computing {\tt posSecond}.

It is clear that if the respective constraints are satisfied for all choices of {\tt start} and all possible labels, then the word that satisfies our model avoids {\tt patterns[p]}  for all choices of {\tt p} successfully.  
For instance, the following  generated word of length 50  over 3 letters  avoids Zimin-3 ( x1x2x1x3x1x2x1): {\tt [1, 1, 1, 1, 1, 1, 2, 1, 1, 2, 1, 1, 2, 2, 1, 1, 3, 1, 1, 2, 2, 2, 1, 1, 3, 2, 1, 1, 3, 2, 1, 1, 3, 3, 1, 1, 2, 2, 3, 1, 1, 2, 2, 3, 1, 1, 2, 3, 1, 1]}
Finished in 57m 14s. This input file  generates by calling {\tt java InputGenerator t 2 25 25 3 input x1x2x1x3x1x2x1}, meaning that considering a prefix of length 2 of the Thue Morse word, each letter mapped to some string of length 25, so we get 50 arbitrary letters.
As another example, the word {\tt [1, 2, 3, 1, 4, 1, 4, 1, 2, 3, 1, 4, 1, 2, 3, 1, 2, 3, 1, 4, 1, 4, 1, 2, 3, 1, 2, 3, 1, 4]}, generated by our model, does not have instances of the pattern {\tt x1x2x2x1r} over an alphabet of size 4. This word of length $30$ was obtained in less than one second on a standard desktop computer. Clearly, this pattern is already avoidable over three letter alphabets, for instance by the Hall word.
\subsubsection{Checking the Avoidability of Patterns in the Abelian Sense.} 
In our second example, we design a model that is satisfied by a word that avoids a certain formula in the Abelian sense.
As explained, we first generate an input file for the MiniZinc model. We use the program {\tt InputGenerator.java} with arguments {\tt t/h} {\tt morphicWordLength} {\tt morphicWordImageLength} {\tt sigma} {\tt sigma} {\tt inputFileName} {\tt pattern1} \ldots.  For example, a list of such arguments is:  {\tt t 5 2 3 3 input xxx}.

The only main difference with respect to the code above occurs when we have computed the length of the image of a pattern {\tt patterns[p]} from the formula, and we are in the case when this image fits in the suffix of {\tt word} that starts on {\tt start}. Now, as above, in the string occurring at position {\tt start} in {\tt word}, whose length equals the length of the image {\tt patterns[p]}, we want to identify two factors corresponding to the image of the same variable, which are not equivalent in the Abelian sense, or in other words do not have the same Parikh vector. Basically, our constraints asks for the existence of a letter (occurring on position {\tt l} in the string that is supposed to be the image of the variable $x$) whose number of occurrences in the string corresponding to $x$ starting on position {\tt occInWord} is not equal to its number of occurrences in the string corresponding to $x$ starting on position {\tt nextOccInWord}. To achieve this we count how many times this letter occurs in the each of the two strings that should correspond to $x$ by summing up its occurrences in these images, respectively. Then we ask that these two sums are not equal. 

This is entire strategy implemented following the main ideas of the previous section, as shown in Listing~6, with the constraint on line 22 changed as described in Listing~7. 

\begin{lstlisting}[caption=Constraints for abelian avoidability]
sum(k in occInWord..(occInWord + length(patterns[p, varOcc], wordLength, nrVarsInPattern[p], i) - 1) where word[k] == word[firstPos]) (word[k]) != 
sum(k in nextOccInWord..(nextOccInWord + length(patterns[p, nextOcc], wordLength, nrVarsInPattern[p], i) - 1) where word[k] == word[firstPos]) (word[k])

\end{lstlisting}
For example, the word {\tt[1, 2, 1, 3, 1, 2, 1, 4, 1, 3, 2, 3, 4, 1, 4, 2, 3, 2, 1, 4, 1, 2, 1, 3, 1, 2, 1, 4, 2, 3, 2, 1, 4]}, obtained with our model, does not have abelian instances of the pattern {\tt xx} over an alphabet of size 5. Considering a prefix of length 5 of the Hall word, each letter mapped to some string of length 8, 7 and 5, so we get 33 arbitrary letters Again. The running time on a standard desktop computer  was 18m 40s.

\subsubsection{Checking Avoidability of Patterns under Permutations.} In this section we address the avoidability of patterns under permutations. Recall that a pattern under permutations, of length $m$, is a pattern  $\pi_1^{i_1}(x_1)\pi_2^{i_2}(x_2)\cdots \pi_m^{i_m}(x_m)$, where, for $1\leq i\leq m$, $x_i$ is a word variables and $\pi_i$ is a functional variable, to be replaced by morphic or antimorphic permutation of the alphabet of terminals (see \cite{unaryJames}). For example, $x \pi(y) \pi(\pi(x)) \sigma(y)$ is a pattern under permutations, where $x,y$ a are word variables and $\pi$ and $\sigma$ are functional variables which can be replaced by morphic or antimorphic permutations of the alphabet of terminals. Note that, for simplicity of the exposure, we exclude the case when multiple different functions are applied on the same word variable, i.e, we exclude cases like $\pi(\sigma(x))$. 

Now, given a formula consisting of patterns under permutations (i.e., a set of patterns under permutations) and an alphabet $\Sigma_k$, we want to design a model that is satisfiable if and only if there exists a word, whose length is also given as input, which avoids the respective formula over $\Sigma_k$. Just like before, the model gets as input a data file, which is constructed automatically by a Java program from a formula that is given by the user. Basically, the user is required to run the Java program {\tt InputGenerator.java} with the input {\tt sigma}, the size of the alphabet, {\tt wordLength}, the length of the word to be generated, the name of the data file to be generated (say {\tt Input.dzn}), and the actual formula written as a sequence of patterns, where the word variables are symbols from the set $\{x_1,x_2,x_3,\ldots\}$ and the function variables are symbols from the set $\{p_1,p_2,\ldots\}$. We allow reversed variables, denoted $x_ir$, and antimorphic permutations denoted as $p^k_i(x_j)r$; in the latter, we encode that an antimoprhism $p_i$, iterated $k$ times, is applied to the word variable $x$. Let us exemplify this.  Consider the pattern $x_1x_2 p^5_1(x_2) p^3_2(x_1)$ where $p_1$ is to be replaced by morphic permutations and $p_2$ by antimorphic permutations. This will be given as parameter to the Java program as {\tt x1x2p1$\hat{\ }$5(x2)p2$\hat{\ }$3(x1)r}. 

Now, for a set of patterns the program {\tt InputGenerator.java} generates the data file {\tt Input.dzn}. This file is now more complex. It contains several simple numerical values {\tt  sigma, wordLength, nrPatterns} with the same meaning as before. Moreover, we compute and store the integer {\tt nrPermutations} which is simply the total number of permutations over an alphabet of size {\tt sigma}, so {\tt sigma!}. Also, we set {\tt maxNrOccs} as the maximum length of a pattern, i.e. the maximum number of items (all occurrences -not necessarily distinct- of word variables or word variables under morphic or antimorphic permutations) occurring in a pattern, just as we did before. The more complex part is how to encode a pattern. Essentially, a pattern under permutations, of length $k$, will be encoded as a sequence of $k$ $4$-tuples as follows. The $i^{th}$ word variable occurring in the pattern (when read left to right) is mapped to the number $i$; also, the $i^{th}$ functional variable occurring in the pattern is mapped to the number $i$. Now, if on position $i$ of the pattern we have $p^k(x)$ (so $p$ is morphic) where $x$ is mapped to $i$ and $p$ to $j$, we encode this as $(j,k,i,0)$; if on position $i$ of the pattern we have $p^k(x)r$ (so $p$ is antimorphic) where $x$ is mapped to $i$ and $p$ to $j$, we encode this as $(j,k,i,1)$. If  on position $i$ of the pattern we have $x$ where $x$ is mapped to $i$, we encode this as $(1,0,i,0)$ (respectively, $(1,0,i,1)$ if we would have had $xr$ on the $i^{th}$ position). It is worth emphasising that we see word variables on which no function variable is applied as word variables on which we apply the identity morphism, so $p^0$, where $p$ is the first morphism occurring in the pattern. An example is given in the following listing. These tuples are kept in a $3$-dimensional array {\tt repetitions}. 

\begin{lstlisting}[caption=Encoding of the pattern $x_1x_2p_2(x_1)$]
  1,0,1,0,    1,0,2,0,    1,1,1,0,    
\end{lstlisting}

The next important part is how to encode the permutations in the file. We will use a $4$-dimensional array {\tt permutations}, where {\tt permutations[i][j]} gives us all the possible ways in which the $j^{th}$ permutation acts each time it occurs in the pattern (that is, how $p^k$ is defined, each time some $p^k$ appears in the pattern, where $p$ is the $j^{th}$ permutation). In the following listing we have {\tt permutations[1][1]} for the pattern $x_1x_2p_2(x_1)$. Note that here $p_2$ is actually the first permutation occurring in the pattern (so, the name the user uses is not important, as it is rehashed to the correct number by our program). 

\begin{lstlisting}[caption={\tt permutations{[1][1]}}]
  (1,2,3),  (1,2,3),  (1,2,3),    
  (1,2,3),  (1,2,3),  (1,3,2),    
  (1,2,3),  (1,2,3),  (2,1,3),  
  (1,2,3),  (1,2,3),  (2,3,1),  
  (1,2,3),  (1,2,3),  (3,1,2),  
  (1,2,3),  (1,2,3),  (3,2,1)    
\end{lstlisting}

Finally, we add to {\tt Input.dzn} two one-dimensional arrays {\tt nrVarsInPattern} and, respectively, {\tt nrPermsInPattern} which simply encode for each pattern of the formula the number of variables, respectively, functional variables occurring the respective pattern. 

Now we can describe how the MiniZinc model works, given the {\tt Input.dzn} data file. Like in the case of a formula with reversals, for each pattern of the formula we check separately whether the generated word contains an image of it. The check is done mainly just like in the case of formulas with reversals: we assign possible lengths to the word variables, and then we check if there exists an assignment of this word variables, as well as one assignment of the function variables that make them fit a factor of the generated word. The main difference is done in the actual check, which is performed as follows. Firstly, for the pattern with the index {\tt p} in the set of patterns we identify the occurrences of the same variable in the pattern, with permutations applied on it.
 
\begin{lstlisting}[caption=Checking an occurrence of a pattern under permutations - part $1$]
exists(m in 1..nrPermsInPattern[p]) (
    forall(z in 1..nrPermutations) (        
        exists(varOcc in 1..maxPatternLength where
            repetitions[p, varOcc, 3] != 0 /\ repetitions[p, varOcc, 1] == m) (
                exists (nextOcc in 1..maxPatternLength where
                      (nextOcc != varOcc /\
                      (repetitions[p, varOcc, 3] == repetitions[p, nextOcc, 3]) /\
                      (repetitions[p, nextOcc, 2] == 0 \/
                       repetitions[p, varOcc, 1] == repetitions[p, nextOcc, 1]))) )))
         % actual check will be performed here  - see part $2$
\end{lstlisting}
More precisely, we need a certain position of $p$, where a variable actually occurs. As such we look for the position {\tt varOcc}, with {\tt repetitions[p, varOcc, 3] != 0}. This means that on the certain position we really have a variable (as such or under a permutation) in the current pattern. Such a check is needed because in the case of multiple patterns, when one is shorter it may contain tuples which consist of $0$s. See the following example.
\begin{lstlisting}[caption=The encoding of the patterns {\tt x1x2p2(x1)} and {\tt x1p1(x1)}]
repetitions = array3d(1..numberOfPatterns, 1..maxNumberOfRepetitions, 1..4, [
  1,0,1,0,    1,0,2,0,    1,1,1,0,    
  1,0,1,0,    1,1,1,0,    0,0,0,0,]);
\end{lstlisting}

Then, for the variable occurring on {\tt varOcc} in the considered pattern we find its next occurrence on position {\tt nextOcc} (i.e., {\tt repetitions [p, varOcc, 3] == repetitions [p, nextOcc, 3])}) under the same permutation (so we have {\tt repetitions[p,varOcc,1]= repetitions[p,nextOcc,1]}) or without any permutation applied on it (i.e., {\tt repetitions [p, nextOcc, 2] == 0}).

%

Then based on a similar method as in the model developed for formulas with reversals, but including now the usage of morphic or antimorphic permutations (lines 10 for the morphic case and, respectively, line 11 for the antimorphic case, from Listing~11) , we check whether there exists a variable in the pattern occurring multiple times (identified as above), each of its occurrences being under a permutations, whose occurrences are mapped to words which are correctly mapped by the corresponding morphic or antimorphic permutations.  
  The following MiniZinc code is doing this check, corresponding to the lines 7-23 from Listing~6.

\begin{lstlisting}[caption=Checking an occurrence of a pattern under permutations - part $2$]
exists(l in 1..varLength) (
    let { var int: occInWord = start + (if varOcc > 1 then
            (sum(k in 1..(varOcc - 1))(length(repetitions[p, k, 3], wordLength, nrVarsInPattern[p], i))) else 0 endif),
        var int: nextOccInWord = start + (if nextOcc > 1 then
            (sum(k in 1..(nextOcc - 1))(length(repetitions[p, k, 3], wordLength, nrVarsInPattern[p], i))) else 0 endif),                
        var int: posFirst = occInWord + l - 1,            
        var int: posSecond =
            (if (repetitions[p, varOcc, 4] == repetitions[p, nextOcc, 4]) then
                (nextOccInWord + l - 1)
                else (nextOccInWord + varLength - (l - 1) - 1) endif)
    } in
    permutations[p, z, varOcc, word[posFirst]] != permutations[p, z, nextOcc, word[posSecond]])
\end{lstlisting}
For example, the  word of length 54 {\tt[1, 2, 3, 1, 4, 2, 1, 4, 3, 2, 4, 3, 1, 4, 3, 2, 1, 3, 2, 4, 1, 2, 4, 3, 1, 2, 3, 1, 4, 2, 1, 4, 3, 2, 1, 3, 2, 4, 1, 2, 4, 3, 1, 4, 3, 2, 4, 3, 1, 2, 3, 1, 4, 2]} obtained with the model described above does not have instances of the patterns {\tt x1x1, x1p1(x1)x1r} over an alphabet of size 4. The input file for this example can be generated by calling  {\tt java InputGenerator h 7 6 6 12 4 input x1x1 x1p1(x1)x1r},  meaning that considering a prefix of length 4 of the Hall word, each letter mapped to some string of length 6, 6 and 12, so we get 54 arbitrary letters. As you can see this is the same word obtained in \cite{unaryMike} Lemma 9. Also in this case this word was obtained in 5m 6s. As another example of this paper the pattern $xp^5(x)p^{12}(x)$ is UNSATISFIABLE or unavoidable on $\Sigma_2$, and the word {\tt[1, 1, 2, 2, 1, 3, 2, 1, 3, 2, 2, 3, 1, 2, 2, 1, 3, 3, 2, 2, 3, 1, 2, 2, 1, 3, 3, 1, 1, 2, 2, 1, 3, 2, 1, 3]} that is an image of the Thue Morse word over 3 letters, does not contain an instance of such pattern.  

\subsubsection{Generating Morphic Words.}\label{morphic-words} 
Generating long words that avoid a certain pattern or formula is usually just a first step in showing avoidability results. In many cases, one is interested in generating such a word that is the morphic image of a word whose structure is well known and studied. To this end, we enhanced our models with additional constraints so that the generated words are also the image of prefixes of well understood infinite words; we only did this for the Thue-Morse word and for the Hall word (also known as the ternary Thue-Morse word). We restrict our search for morphisms that map the letters of the {two aforementioned words {\tt t} or {\tt h} to words of given length. 

{

Let us describe our approach in more details.
Firstly,  we generate an input file for our model by giving several arguments to a Java program {\tt InputGenerator}.  The first argument is either {\tt t} or {\tt h} and specifies which initial infinite word we use: the binary Thue word or, respectively, the Hall word. This word is called {\tt morhpicWord} in the code. Then we give the length that this initial word should have. Then, as a list of integers, the lengths of the images of each letter of the initial word ({\tt t} or {\tt h}) under the morphism that will define the word our model generates. For {\tt t} we need to give these lengths for the letters $0$ and $1$, while for the Hall word we need to give the lengths for the letters $0, 1, 2$. The rest of the arguments are given as before.
The next example contains a section of an input file used to generate the final morphic words.
\begin{lstlisting}[caption={Specification for the word {\tt t} and the way it should be mapped.}]
morphicWordLength = 10;
morphicWord = array1d(1..morphicWordLength, [  0, 1, 1, 0, 1, 0, 0, 1, 1, 0, ]);
numberOfMorphicWordImages = 2;
morphicWordImagesLengths = array1d(1..numberOfMorphicWordImages, [  2, 3, ]);
\end{lstlisting}
Note that here {\tt morphicWordImagesLengths} is an array that specifies the lengths of the images of the letters of the morphic word under the morphism that will map it to the word avoiding the given patterns. In this example, $2$ is the length of the image of letter $0$, and $3$ is the length of the image of $1$ for the word {\tt t}. 

Once the input file is constructed, we can proceed and describe how the final word is constructed by the Minizinc model. 
The main idea is to construct a template-word, which has the desired avoidability properties, whose length is the sum of the lengths of the images of the letters of the morphic word. This word is used as a template for the actual images of the letters of the morphic word. Its first $morphicWordLength[1]$ letters are the image of $0$, the next $morphicWordLength[2]$ letters are the image of $1$, and so on. This word is called {\tt word}. We also construct {\tt finalWord}, which is supposed to be the word that has the desired avoidability properties and is the morphic image of {\tt morphicWord}, under the morphism defined with the help of {\tt word}. The two words described above are defined as in the following listing.
\begin{lstlisting}[caption={\tt {}}]
constraint avoidPatterns(word, wordLength);
constraint avoidPatterns(finalWord, (sum(i in 1..morphicWordLength) 
												        	(morphicWordImagesLengths	[morphicWord[i] + 1])));
\end{lstlisting}

The only thing left to do is to check that {\tt finalWord} is the image of {\tt morphicWord}, by the morphism defined using the factors of {\tt word}. This is done in the following.
\begin{lstlisting}[caption=Checking whether {\tt finalWord} is a morphic image]
constraint
  forall (i in 1..morphicWordLength)(
    let { var int: morphCharPos = morphicWord[i] + 1;
      var int: posInWord = (sum (k in 1..(morphCharPos - 1)) (morphicWordImagesLengths[k])) + 1,
      var int: posInFinalWord = (sum (k in 1..(i - 1)) (morphicWordImagesLengths[morphicWord[k] + 1])) + 1,
    } in
    forall (j in 1..morphicWordImagesLengths[morphCharPos])(
      word[posInWord + j - 1] = finalWord[posInFinalWord + j - 1]));
\end{lstlisting}

Here {\tt morphCharPos} gives the current letter of {\tt morphicWord}. Then {\tt posInWord} tells us where the image of this letter starts in {\tt word}. Finally, {\tt posInFinalWord} gives the current position in {\tt finalWord}. We then just have to check whether the  factor of length {\tt morphicWordImagesLengths[morphCharPos]} occurring in {\tt word} at position {\tt posInWord} is the same as the one of equal length occurring in {\tt finalWord} at {\tt posInFinalWord}. If this last constraint is satisfied, then {\tt finalWord} is the morphic image of {\tt morphicWord}, and has the required avoidabiity properties.

}

For example, the word {\tt [1, 1, 2, 2, 3, 2, 3, 1, 1, 2, 2, 3]} is generated by MiniZinc as the morphic image of the Thue Morse word of length 5, i.e., {\tt 0, 1, 1, 0, 1}, such that the length of the image of 0 is equal to 3, i.e., {\tt 112}, and the length of the image of 1 is equal to 2, i.e., {\tt 23}. 

\newpage

\section*{Appendix}
\subsection{The MiniZinc Language}
In the following, we give a short overview of how MiniZinc works. For more details, see \url{http://www.minizinc.org/downloads/doc-latest/minizinc-tute.pdf}. 

Information for the reference and use of programs is stored by employing variables or parameters, which are declared and assigned a type, which, at its turn, gives them their value. The fundamental types of parameters are strings (string), integers (int), Booleans (bool) and floating point numbers (float). MiniZinc also supports arrays and sets. As such, one-and multi-dimensional arrays, declared as {\tt array[<index-set1>, \dots, <index-setn>~] of <type-inst>}. 
MiniZinc has a requirement for the array declaration to contain the index set of each dimension. This index set must either be a set variable initialized to an integer range, or an integer range itself (as we use here). Arrays may hold any of the base types.
In our models, we also use the {\tt sum} function which provides the arithmetic arrays aggregation function which adds its element.

MiniZinc models may also employ another type of variables, namely decision variables, which are variables in the logical sense. They differ from variables and parameters from standard programming languages in that there is no need for the model to assign them a value. Their values remain unknown, until, during the execution of the MiniZinc model, the solving system decides that it is possible for a certain decision variable to be given a certain value satisfying the model's constraints.
MiniZinc makes a careful distinction between parameter and decision variables. 

As far as the syntax is concerned, variables are assigned values by {\em assignment items}. They take the form: {\tt <variable>=<expression>;}.  The most important part of a model are the {\em constraint items}. These take the form: 
{\tt constraint< Boolean~exp-\\ression>;}. {\tt Forall} and {\tt exists} conditions can be used for arrays of constraints: {\tt forall} ensures that every constraint in an array holds, while {\tt exists} ensures that at least one constraint holds. 
{\em Solve items} define precisely the type of solution being sought in our model. In our case we will use only {\tt solve satisfy;} items.  In this instance, the problem is a constraint satisfaction problem: we need to discover a value for the decision variables that can satisfy the constraints; the exact value is not important (as it would be in the case of optimisation problems). The last element of the model is the {\em output statement}. This statement informs MiniZinc what it should print once the model has run and a solution has been discovered. Output items give a good presentation of the model execution's results. They take the form:
{\tt output [~<string~expression>,\dots, <string~expression>~];}.

The solutions we propose are based on the following standard workflow.
Firstly, we generate the {\em data files}, which encode the input to our model. Secondly, {\em compile the MiniZinc model together with the data file} into Flatzinc. Thirdly, in an implicit step, run a CSP solver on the Flatzinc file.

The generation of the data files is done by executing a Java program. Depending on the problem we solve, we need to proceed at this step as follows (the precise semantic of each parameter is explained in the respective sections). 

The main codes for this paper are in the \texttt{Morphic} directory, however, for those who may be interested in generating simple words which are not images of the morphisms, we have codes in \texttt{None Morphic} directory. They were anyway written as an initial step of our approach. 

We give the following parameters as the input arguments of the programs sInputGenerator.java: {\tt t/h morphicWordLength morphicWordIma-\\gesLengths sigma inputFileName pattern1 [pattern2 ...].}

These  will generate .dzn data files, with the name {\tt inputFileName}. The next steps consist in running MiniZinc and the CSP solver. In our case, one can either open the MiniZinc model from the folder minizinc, using a standard IDE, or call the tools from the command line:
 \begin{itemize}
\item[-]  cd minizinc
\item[-] mzn2fzn rep.mzn ../permGen/Input.dzn (generates the flatzinc file)
\item[-] fzn-gecode rep.fzn (runs gecode on this example)
\end{itemize}
To set the value of parameters using the generated .dzn data files, on the configuration part of the MiniZinc, and on the Data file menu, one needs to choose the specific data file that you want to use.  

We verified that the results generated using Minizinc are correct solutions to the avoidability problems we considered using a checker program ResultChecker.java. This Java program is using a standard backtracking algorithm to check whether the word which is obtained via Minizinc contains instances of the patterns that we wanted to avoid or not. The result was that the generated words did not contain instances, so the Minizinc model produced a correct solution. The input arguments for the java program should be: {\tt solution sigma wordLength pattern1 [pattern2 pattern3 ...]}. Note that we should give the solution generated by Minizinc model between quotation marks as the {\tt solution} argument of the java program. An example of possible arguments for this program are: {\tt "[1, 2, 3, 1, 4, 1, 4, 1, 2, 3]" 5  10 x1x2x2x1r}. 
\\
\\
\textbf{Longer words generate by MiniZiinc}
\\
\\
This is the generated word by our model (considering that $x1$ and $x2$ and in general all variable are always none empty) for sigma = 3; wordLength = 120; pattern = x1x2x2x1r,
 morphicWord = a,b, morphicWordImagesLengths=60,60:\newline
[1, 1, 1, 2, 1, 1, 1, 2, 1, 2, 2, 2, 1, 1, 1, 2, 1, 1, 1, 3, 1, 1, 1, 2, 1, 1, 1, 2, 1, 2, 2, 2, 1, 1, 1, 2, 1, 1, 1, 3, 1, 2, 1, 1, 1, 2, 1, 1, 1, 2, 1, 2, 2, 2, 1, 1, 1, 2, 1, 1, 1, 3, 1, 1, 1, 2, 1, 1, 1, 2, 1, 2, 2, 2, 1, 1, 1, 2, 1, 1, 1, 3, 1, 2, 1, 2, 1, 2, 1, 2, 1, 2, 1, 2, 1, 2, 1, 2, 1, 2, 1, 2, 1, 2, 1, 2, 1, 2, 1, 2, 1, 2, 1, 2, 1, 2, 1, 2, 1, 2]
Finished in 1h 1m 37s
\\
\\
Now, we want to compare the running time of  MiniZinc  with a java code using a similar algorithm. Below you can find run time comparison for avoidability of the pattern x1x2x2x1r mentioned above over an alphabet of size 5. To generate words by a naive backtracking algorithm using Java, we use ResultChecker.java file exists in the \texttt{None morphic} folder. The input arguments for this code to generate can be: \texttt{Unsatisfiable 5 120 x1x2x2x1r}. Note that, this Unsatisfiable option is only available for none morphic words, in order to compare the running times.

\begin{table}[]
  \centering
\begin{tabular}{cccll}
\hline
\multicolumn{4}{c}{MiniZinc}                                                                                       & \multicolumn{1}{c}{Java}      \\
\hline
\multicolumn{1}{l}{WordLength} & \multicolumn{1}{l}{SolveTime(s)} & \multicolumn{1}{l}{InitTime(s)} & Finished(s) & \multicolumn{1}{c} {RunTime(s)} \\
80                             & 0.121                            & 2.355                           &        \multicolumn{1}{c}{192}     & \multicolumn{1}{c}{2.039 }  \\
90                             & 0.251                            & 3.648                           &      \multicolumn{1}{c}{294}      & \multicolumn{1}{c}{3.052 }  \\
100                            & 0.329                            & 5.569                           &         \multicolumn{1}{c}{433}     &\multicolumn{1}{c}{ 4.031 }     \\
110                            & 0.503                            & 8.028                           &           \multicolumn{1}{c}{607}  & \multicolumn{1}{c}{13.609}     \\
120                            & 0.779                            & 11.348                          & \multicolumn{1}{c}{860} & \multicolumn{1}{c}{18.183}       \\
\hline
\end{tabular}
\end{table}

In the end, we mention some factors which may affect the MiniZinc performance:
 \begin{itemize}
\item There is a really easy (easy to find) solution. In this case, the Minizinc-toolchain will generate (i.e., compile) a large formula which considers many constraints, taking lots of time for this. Afterward, solving is rather fast. However, a naive backtracking may run even faster, especially if you consider solving time + compilation, because you hardly have to backtrack, if ever.
 \item The formula is unsatisfiable, i.e. the desired word does not exist. Then, naive backtracking is extremely slow, the worst case being a generate-and-check approach, which might generate $|\Sigma|^n$ words of length n, and decide that none of them satisfies the given constraints. In this case, smarter backtracking, or constraint programming, should be way faster, in many cases exponentially faster. The compilation time may still be high, but it is worth the extra time because solving is fast afterward.
 \item There is a solution which is tricky to find. This is a complicated case - given the right search strategy, backtracking may be extremely fast (guess the right solution, and just check that it is correct). However, you cannot rely on making a good guess. Thus, in most cases, your initial guess will be somewhat wrong, and the time you pay for compilation makes sense because solving is faster in most cases. However, there may be cases in which backtracking is faster. On the other hand, constraint solving should provide a   solution even if the initial guess is bad, while backtracking may be stuck in the wrong part of the search space for ages.
 \end{itemize}
  
This work is the first attempt, that we are aware of, to solve avoidability problems as CSPs, and it would be interesting to continue this initial steps .  It opens the door to a new research direction,  and  there is space for improvements

\end{document}